\documentclass[nofootinbib, preprint, aps, prc, superscriptaddress,11pt]{revtex4-1}
\usepackage[utf8]{inputenc}

\usepackage{epsfig}
\usepackage[utf8]{inputenc}
\usepackage{amsmath}
\usepackage{amssymb}
\usepackage{amsthm}
\usepackage{amsfonts}

\usepackage{xspace}
\usepackage{graphicx}
\usepackage{graphics}
\usepackage{slashed}
\usepackage{bbold}
\usepackage{bbm}
\usepackage{empheq}
\usepackage{mathrsfs}
\usepackage{xcolor}
\usepackage{slashed}

\usepackage{setspace} 
\usepackage{comment}

\usepackage{cancel}
\usepackage{mathtools}

\usepackage{hyperref}

\usepackage{siunitx}
\allowdisplaybreaks

\newcommand{\Nc}{\ensuremath{N_c}\xspace}

\newcommand{\NN}{\ensuremath{N\!N}\xspace}

\newcommand{\oneS}{\ensuremath{{{}^{1}\!S_0}}\xspace}
\newcommand{\threeS}{\ensuremath{{{}^{3}\!S_1}}\xspace}
\newcommand{\fiveS}{\ensuremath{{{}^{5}\!S_2}}\xspace}
\newcommand{\sevenS}{\ensuremath{{{}^{7}\!S_3}}\xspace}

\newcommand{\threeD}{\ensuremath{{{}^{3}\!D_1}}\xspace}

\newcommand\Swave{$S$-wave\xspace}
\newcommand\Pwave{$P$-wave\xspace}
\newcommand\Dwave{$D$-wave\xspace}
\newcommand\SD{\ensuremath{\threeS\text{-}\threeD}\xspace}
\newcommand\SP{\ensuremath{S\text{-}P}\xspace}

\newcommand{\calO}{\ensuremath{\mathcal{O}}}

\newcommand{\calA}{\ensuremath{\mathcal{A}}}
\newcommand{\calL}{\ensuremath{\mathcal{L}}}

\newcommand{\Czerotrip}{\ensuremath{C_{\NN,\slashed{\D}}^{(^3 \! S_1)}}\xspace}
\newcommand{\Czerosing}{\ensuremath{C_{\NN,\slashed{\D}}^{(^1 \! S_0)}}\xspace}

\newcommand{\ConeP}{\ensuremath{C^{(^1 \! P_1)}}\xspace}

\newcommand{\CPone}{\ensuremath{C^{(^3 \! P_1)}}\xspace}
\newcommand{\CPtwo}{\ensuremath{C^{(^3 \! P_2)}}\xspace}

\newcommand{\CDN}{\ensuremath{C_{\Delta N}}\xspace}

\newcommand{\CNNsing}{\ensuremath{C_{\NN}^{(^1 \! S_0)}}\xspace}
\newcommand{\CNDsing}{\ensuremath{C_{\Delta N}^{(^1 \! S_0)}}\xspace}
\newcommand{\CDDsing}{\ensuremath{C_{\Delta\Delta}^{(0,1)}}\xspace}
\newcommand{\ONNsing}{\ensuremath{\calO_{\NN}^{(^1 \! S_0)}}\xspace}
\newcommand{\ONDsing}{\ensuremath{\calO_{\Delta N}^{(^1 \! S_0)}}\xspace}

\newcommand{\CNNtrip}{\ensuremath{C_{\NN}^{(^3 \! S_1)}}\xspace}
\newcommand{\CNDtrip}{\ensuremath{C_{\Delta N}^{(^3 \! S_1)}}\xspace}
\newcommand{\CDDtrip}{\ensuremath{C_{\Delta\Delta}^{(1,0)}}\xspace}
\newcommand{\ONNtrip}{\ensuremath{\calO_{\NN}^{(^3 \! S_1)}}\xspace}
\newcommand{\ONDtrip}{\ensuremath{\calO_{\Delta N}^{(^3 \! S_1)}}\xspace}

\newcommand{\CeffSing}{\ensuremath{\widetilde C_{\NN}^{(\oneS)}}\xspace}
\newcommand{\CeffTrip}{\ensuremath{\widetilde C_{\NN}^{(\threeS)}}\xspace}
\newcommand{\Ceff}{\ensuremath{\widetilde C_{\NN}^{(S)}}\xspace}
\newcommand{\Ceffprime}{\ensuremath{\widetilde C_{\NN}^{(S')}}\xspace}

\newcommand{\CsdNN}{\ensuremath{C^{(\SD)}_{\NN}}\xspace}
\newcommand{\CsdNDelta}{\ensuremath{C^{(\SD)}_{\D N}}\xspace}
\newcommand{\CsdDless}{\ensuremath{C^{(\SD)}_{\slashed{\D}}}\xspace}
\newcommand{\Ceffsd}{\ensuremath{\widetilde C_{\NN}^{(\SD)}}\xspace}

\newcommand{\CPVNN}{\ensuremath{C^{(\SP)}_{\NN}}\xspace}
\newcommand{\CPVNDelta}{\ensuremath{C^{(\SP)}_{\D N}}\xspace}
\newcommand{\CPVDless}{\ensuremath{C^{(\SP)}_{\slashed{\D}}}\xspace}
\newcommand{\CPVDlessprime}{\ensuremath{C^{(S'\text{-}P')}_{\slashed{\D}}}\xspace}
\newcommand{\CPVeff}{\ensuremath{\widetilde C_{\NN}^{(\SP)}}\xspace}
\newcommand{\CPVeffprime}{\ensuremath{\widetilde C_{\NN}^{(S'\text{-}P')}}\xspace}

\newcommand{\at}{\ensuremath{\tilde{a}}\xspace}
\newcommand{\bt}{\ensuremath{\tilde{b}}\xspace}

\newcommand{\CDDoneSIone}{C_{\DD}^{(0,1)}}
\newcommand{\CDDoneSIthree}{C_{\DD}^{(0,3)}}
\newcommand{\CDDthreeSIzero}{C_{\DD}^{(1,0)}}
\newcommand{\CDDthreeSItwo}{C_{\DD}^{(1,2)}}
\newcommand{\CDDfiveSIone}{C_{\DD}^{(2,1)}}
\newcommand{\CDDfiveSIthree}{C_{\DD}^{(2,3)}}
\newcommand{\CDDsevenSIzero}{C_{\DD}^{(3,0)}}
\newcommand{\CDDsevenSItwo}{C_{\DD}^{(3,2)}}

\newcommand{\CDDeffSing}{\ensuremath{\widetilde C_{\Delta\Delta}^{(0,1)}}\xspace}

\newcommand{\CDDeffTrip}{\ensuremath{\widetilde C_{\Delta\Delta}^{(1,0)}}\xspace}

\newcommand{\COOoneS}{C_{\OO}^{(\oneS)}}
\newcommand{\COOfiveS}{C_{\OO}^{(\fiveS)}}

\newcommand{\asing}{\ensuremath{a_{\NN}^{(^1 \! S_0)}}\xspace}

\newcommand{\atrip}{\ensuremath{a_{\NN}^{(^3 \! S_1)}}\xspace}

\newcommand{\aNN}{\ensuremath{a_{\NN}}\xspace}

\newcommand{\aDDthreezero}{\ensuremath{a_3}\xspace}

\newcommand{\aOOsing}{\ensuremath{a_{\OO}^{(\oneS)}}\xspace}

\newcommand{\mpi}{\ensuremath{m_\pi}}

\newcommand{\nopi}{\ensuremath{\pi\hskip-0.40em /}}
\newcommand{\eftnopi}{EFT$_{\nopi}$\xspace}

\newcommand{\D}{\ensuremath{\Delta}\xspace}
\newcommand{\DD}{\ensuremath{\Delta\Delta}\xspace}
\newcommand{\ND}{\ensuremath{N\!\Delta}\xspace}
\newcommand{\DN}{\ensuremath{\Delta N}\xspace}
\newcommand{\Om}{\ensuremath{\Omega}\xspace}
\newcommand{\OO}{\ensuremath{\Omega\Omega}\xspace}

\begin{document}

\title{The role of intermediate \DD states in nucleon-nucleon scattering in the large-\Nc and unitary limits, and \DD and \OO scattering}

\author{Thomas R.~Richardson}
\email{thomas.richardson@berkeley.edu}
\affiliation{Department  of  Physics,  Duke  University,  Durham,  NC  27708,  USA}
\affiliation{Institut f\"ur Kernphysik and PRISMA$^+$ Cluster of Excellence, Johannes Gutenberg-Universit\"at, 55128 Mainz, Germany}
\affiliation{Department of Physics, University of California, Berkeley, CA 94720, USA}
\affiliation{Nuclear Science Division, Lawrence Berkeley National Laboratory, Berkeley, CA 94720, USA}
\author{Matthias R.~Schindler}
\email{mschindl@mailbox.sc.edu}
\affiliation{Department of Physics and Astronomy,
University of South Carolina, 
Columbia, SC 29208, USA}
\author{Roxanne P. Springer}
\email{rps@duke.edu}
\affiliation{Department  of  Physics,  Duke  University,  Durham,  NC  27708,  USA}

\date{\today}

\begin{abstract}

We explore potential explanations for why using large-\Nc (\Nc is the number of colors) scaling to determine the relative size of few-nucleon low-energy operators agrees with experiment even when dynamical \D's are not explicitly included. 
Given that the large-\Nc analysis is predicated on the nucleons and \D's being degenerate, this is a curious result.  
We show that for purely $S$-wave interactions the relationships dictated by large-\Nc scaling are unaffected whether the \D is included or not.
In the case of higher partial waves that do not mix with $S$-waves, the impact of the  \D is perturbative, which makes the  agreement with  naive (\D-less) large-\Nc ordering unsurprising. 
For higher partial waves that mix with $S$-waves, the nucleon and \D would need to decouple to get agreement with naive large-\Nc ordering.
We find all \NN, \DN, and \DD low energy coefficients for leading-order baryon-baryon scattering in \D-full pionless effective field theory in terms of the two independent parameters dictated by the SU($2F$) spin-flavor symmetry that arises in the \Nc $\rightarrow \infty$ limit.  
Because of recent lattice QCD results and experimental interest, we extend our analysis to the three-flavor case to study $\OO$ scattering.
We show that in the unitary limit (where scattering lengths become infinite) one of the two SU($2F$) parameters is driven to zero, resulting in enhanced symmetries, which agree with those found in spin-1/2 entanglement studies.

\end{abstract}

\maketitle

\section{Introduction}

One way to connect interactions between nucleons to the underlying theory of quantum chromodynamics (QCD) is to use effective field theories (EFTs). 
The EFTs are often cast as expansions in small quantities, such as low momenta and, if pions are included as degrees of freedom, the light quark masses, divided by the breakdown scale of the EFT.
The utility of the EFTs can be augmented by considering the consequences of the limiting cases of approximate symmetries of nature.  
Such an approximate symmetry emerges for the baryon spectrum in the large-\Nc limit of QCD \cite{tHooft:1973jz,Gervais:1983wq,Gervais:1984rc,Dashen:1993as}, providing another approach to capture additional nonperturbative aspects of the strong interactions. 
When combined with EFTs, the large-\Nc analysis can establish relationships among the low-energy coefficients (LECs) of the EFT.
For example, the large-\Nc expansion has been applied to the meson and single-baryon sectors of chiral perturbation theory (see, e.g., Refs.~\cite{Moussallam:1994xp,Jenkins:1995gc,Herrera-Siklody:1996tqr,Leutwyler:1997yr,Kaiser:2000gs,Flores-Mendieta:2000ljq,Flores-Mendieta:2006ojy,Flores-Mendieta:2009vss, Flores-Mendieta:2014vaa,Flores-Mendieta:2024kvj,Jayakodige:2025agk}).

Baryon-baryon scattering in the large-\Nc limit was first considered in Ref.~\cite{Witten:1979kh}, in which the momentum was assumed to scale linearly with \Nc. The leading-in-\Nc interaction is $O(\Nc)$, and with the momentum scaling of Ref.~\cite{Witten:1979kh} the kinetic and potential energies are of the same order in the large-\Nc expansion. 
In the following we focus on low-momentum scattering, and we assume that momenta are independent of \Nc.
This scaling follows that chosen by Refs.~\cite{Kaplan:1995yg,Kaplan:1996rk,Banerjee:2001js}.  
In particular, Refs.~\cite{Kaplan:1995yg,Kaplan:1996rk} argue that despite the lack of a smooth large-\Nc limit for the scattering amplitude, the SU(4) symmetry emerging in the baryon sector with this momentum scaling provides relevant constraints on the relative sizes of different spin-isospin contributions to the nucleon-nucleon (\NN) interactions.

The analysis is performed by considering two-nucleon matrix elements of the Hamiltonian,
which in the large-\Nc limit can be expressed in terms of bilinears of quark fields that encode the spin and isospin content \cite{Dashen:1993jt,Dashen:1994qi,Carone:1993dz,Luty:1993fu}.
The resulting 1/\Nc hierarchy of the different spin, isospin, and momentum structures is mapped onto
phenomenological models or effective field theories (EFTs) of the \NN interactions, indicating the relative sizes of the model/EFT parameters that are to be fit to data \cite{Kaplan:1996rk}.
The LECs are typically subtraction point dependent; they are not observables. Nevertheless, at specific subtraction points, LECs can be extracted from data where available. \NN scattering serves as an important test case of establishing the usefulness of the large-\Nc limit for understanding baryon-baryon interactions. However, the main utility of the large-\Nc analysis lies with those cases for which not enough data is available to sufficiently constrain LECs, e.g., symmetry-violating interactions or the coupling to external fields.
For a recent review see Ref.~\cite{Richardson:2022hyj}.

The delta resonances (\D's) might be expected to play a critical role when considering how nucleons ($N$'s) behave in the large-\Nc limit of QCD. 
In particular, the inclusion of the \D's is crucial to obtaining a consistent picture of pion-nucleon scattering \cite{Gervais:1983wq,Gervais:1984rc,Dashen:1993as}. 
As \Nc $\rightarrow \infty$, the baryon spectrum not only exhibits an SU(2$F$) spin-flavor symmetry, where $F$ is the number of flavors, but deviations from this symmetry can be calculated systematically in a perturbative 1/\Nc  expansion \cite{Gervais:1983wq,Gervais:1984rc,Dashen:1993as,Dashen:1993ac,Jenkins:1993af,Dashen:1993jt,Dashen:1994qi}. 
 Restricting the discussion to two flavors, the $N$'s and \D's transform in the same SU(4) multiplet and have degenerate masses, together with a whole tower of states with higher spin and isospin that do not have equivalents in the physical world where $\Nc = 3$. 
This SU(4) symmetry in the baryon sector has been used to derive the large-\Nc scaling of matrix elements of single-nucleon states; see, e.g., Refs.~\cite{Witten:1979kh,Coleman:1985rnk,Manohar:1998xv,Jenkins:1998wy,Lebed:1998st} for reviews.
The scalings of these matrix elements also form the basis of the large-\Nc analysis in the \NN sector.

However, while the large-\Nc analysis (happening at the quark level) implicitly includes the \D, many of the phenomenological models and EFTs do not include \D's as active/explicit  degrees of freedom; instead their effects are subsumed into the parameters of the model or EFT.
That this mismatch of active degrees of freedom might be problematic has been pointed out in Ref.~\cite{Kaplan:1996rk}.
Further, Ref.~\cite{Banerjee:2001js} analyzed the \NN potential in a meson-exchange picture and showed that again the $\Delta$ degrees of freedom are important  for a consistent analysis within this framework.
It is therefore possibly surprising that no disagreement between the large-\Nc expectations and the relative sizes of the model/EFT parameters dictated by data has been found in the two-nucleon sector. 
If the \D is important for theoretical consistency, 
why do the large-\Nc expectations appear to hold in two-nucleon models/EFTs without \D degrees of freedom?

Here we explore this question in the context of pionless effective field theory (\eftnopi) with the inclusion of dynamical \D's. 
In \eftnopi with unnaturally large scattering lengths, $S$-wave interactions are summed to all orders. 
It is not obvious what happens to the scattering lengths in the large-\Nc limit. 
Reference \cite{Beane:2002ab} showed that the deuteron binding energy scales as $B\sim \Nc^{-5}$, corresponding to a scattering length that scales as $\Nc$. 
In this approach a large scattering length is natural in the large-\Nc limit. Reference \cite{Chen:2017tgv} argues for a different overall scaling of the \NN interactions; in this case the \Swave scattering length is independent of \Nc. 
An unnaturally large scattering length in the physical world with $\Nc=3$ then also implies a large \Swave scattering length in the large-\Nc limit.

In what follows we consider a variety of baryon-baryon scattering channels in order to explore the impact of imposing large-\Nc constraints on the \eftnopi with dynamical $\Delta$'s, in addition to analyzing limiting cases that simplify and potentially illuminate our results:  
\begin{itemize}
    \item Section~\ref{sec:background} discusses the SU(4)-invariant Lagrangian of Ref.~\cite{Kaplan:1995yg} that forms the basis of the \Swave  analysis.
    \item Section~\ref{sec:DDIntermediate} describes the inclusion of $\Delta\Delta$ intermediate states in \NN scattering (closely following Ref.~\cite{Savage:1996tb}).
We show that for $S$ waves the large-\Nc expectations  hold for \emph{ratios} of LECs independent of whether the EFT includes dynamical \D's or not. 
    \item Section~\ref{sec:flip} introduces a spin-isospin exchange symmetry that emerges for \Swave interactions. Use of this symmetry greatly simplifies the derivation of the  \Swave results  of the previous sections. 
    \item Section~\ref{sec:Higher} discusses higher partial waves. 
For those that do not couple to $S$ waves, LECs in the \D-less and \D-full theories are expected to be perturbatively close to each other. 
Higher partial waves that mix with $S$ waves, such as \SD mixing and parity-violating interactions that couple $S$ to $P$ waves, fall in a different category. 
In these cases the \NN and \DD sectors need to decouple to maintain the relationships derived from naive large-\Nc scaling.
    \item Section~\ref{sec:BBscattering} considers how different values of the coefficients in the SU(4)  invariant Lagrangian impact the strength of the \NN, $\ND$, and \DD LECs.
We find that significant simplifications arise in the unitary limit, for which the \NN and \DD sectors decouple. 
The proximity of both the \oneS and \threeS \NN scattering lengths to the unitary limit provides another possible explanation for why the large-\Nc analysis can be successfully applied to a \D-less theory.
\end{itemize}

\section{The SU($2F$)-invariant Lagrangian}
\label{sec:background}

Our analysis is based on combining the large-\Nc constraints that can be derived from the four-baryon Lagrangian of Ref.~\cite{Kaplan:1995yg} with the \eftnopi formalism for including intermediate \DD states in \NN scattering of Ref.~\cite{Savage:1996tb}.
In this section we discuss the relevant large-\Nc concepts and notation.
In the large-\Nc limit,  baryon states transform under irreducible representations of a contracted  SU($2F$) spin-flavor symmetry \cite{Gervais:1983wq,Dashen:1993as,Jenkins:1993af,Dashen:1993jt,Dashen:1994qi}, where $F=2,3$ is the number of light quark flavors. 
For the physical value $\Nc = 3$, the ground states of the  lightest baryons form a completely symmetric SU($2F$) multiplet that can be represented by a symmetric tensor 
$\Psi^{\mathcal{A}\mathcal{B}\mathcal{C}}$, where the indices of the  SU($2F$) fundamental representation $\mathcal{A},\mathcal{B},\ldots \in \{1, \ldots, 2F \}$.

The  most general SU($2F$)-invariant, leading-order (zero-derivative) Lagrangian describing  baryon-baryon scattering  is given in terms of $\Psi^{\mathcal{A}\mathcal{B}\mathcal{C}}$ fields by
\cite{Kaplan:1995yg}
\begin{align}
\label{eq:SU4Lag}
    \calL_6 = -\left[ \at\left( \Psi^\dagger_{\mathcal{A}\mathcal{B}\mathcal{C}} \Psi^{\mathcal{A}\mathcal{B}\mathcal{C}}\right)^2 + \bt \Psi^\dagger_{\mathcal{A}\mathcal{B}\mathcal{C}} \Psi^{\mathcal{A}\mathcal{B}\mathcal{D}}\Psi^\dagger_{\mathcal{E}\mathcal{F}\mathcal{D}} \Psi^{\mathcal{E}\mathcal{F}\mathcal{C}}\right],
\end{align}
where $\at$ and $\bt$ are independent LECs.\footnote{These parameters are related to those of Ref.~\cite{Kaplan:1995yg} by $\at = a/f_\pi^2$ and $\bt = b/f_\pi^2$, where $f_\pi =132 \text{ MeV}$ is the pion decay constant.} (The subscript 6 denotes the dimensions of the operators.) This  Lagrangian is also invariant under parity, charge-conjugation, and time-reversal discrete transformations.
Since the terms in Eq.~\eqref{eq:SU4Lag} do not contain any derivatives, the LECs $\at$ and $\bt$ are related to the \NN \Swave LECs \cite{Kaplan:1995yg}. In the following we show how they are also related to other \Swave interactions.
We restrict the discussion to very low energies, such that terms with additional derivatives on the fields are suppressed by powers of $p/\Lambda_b$, where $p$ is the typical momentum and $\Lambda_b$ the breakdown scale of the pionless theory.

\subsection{Particle content of fields}
In the two-flavor ($F=2$) case, the symmetric tensor $\Psi^{\mathcal{A}\mathcal{B}\mathcal{C}}$ is a 20-dimensional representation of SU(4). It is related to the nucleon and \D fields by expressing each SU(4) index in terms of a pair of spin and isospin indices, cf.~\cite{Kaplan:1995yg}, 
\begin{align}\label{symm.field}
    \Psi^{abc}_{\alpha\beta\gamma} = \Delta^{abc}_{\alpha\beta\gamma} + \frac{1}{\sqrt{18}} \left( N^a_\alpha \epsilon^{bc} \epsilon_{\beta\gamma} + N^b_\beta \epsilon^{ac} \epsilon_{\alpha\gamma} + N^c_\gamma \epsilon^{ab} \epsilon_{\alpha\beta}\right)\ .
\end{align}
where the indices in Eq.~\eqref{eq:SU4Lag} correspond to pairs of spin (Greek) and isospin (lowercase Latin) indices in Eq.~\eqref{symm.field}, e.g.,~$\mathcal{A}$ corresponds to the pair $(\alpha,a)$, where $\alpha=1,2$ for spin up and spin down, respectively; and $a=1,2$ for up flavor and down flavor, respectively.
The tensor $\Delta^{abc}_{\alpha\beta\gamma}$ is completely symmetric in its spin indices ($\alpha,\beta,\gamma,\ldots$) and independently  completely symmetric in its isospin indices ($a,b,c,\ldots$).
Its components are related to the physical \D fields by (spin indices suppressed)
\begin{align}
    \D^{111} & = \Delta^{++}\,, & \D^{112} &= \frac{1}{\sqrt{3}}\Delta^{+}\,, & \D^{122} &= \frac{1}{\sqrt{3}}\Delta^{0}\,, & \D^{222} &= \Delta^{-} \, .
\end{align}

For $F=3$, the lowest-lying baryons are in a 56-dimensional representation. 
The completely symmetric tensor $\Psi^{\mathcal{A}\mathcal{B}\mathcal{C}}$ is given by \cite{Kaplan:1995yg}
\begin{align}\label{eq:56plet}
    \Psi^{abc}_{\alpha\beta\gamma} = T^{abc}_{\alpha\beta\gamma} + \frac{1}{\sqrt{18}} \left( B^a_{d,\alpha} \epsilon^{dbc} \epsilon_{\beta\gamma} + B^b_{d,\beta} \epsilon^{dac} \epsilon_{\alpha\gamma} + B^c_{d,\gamma} \epsilon^{dab} \epsilon_{\alpha\beta}\right) \,,
\end{align}
where the flavor indices $a,b,c,d$ now run from 1 to 3. 
$T$ denotes the decuplet fields with (spin indices suppressed)
\begin{equation}
\begin{aligned}
    T^{111} & = \Delta^{++}\,,\quad & T^{112} &= \frac{1}{\sqrt{3}}\Delta^{+}\,, \quad & T^{122} &= \frac{1}{\sqrt{3}}\Delta^{0}\,,\quad & T^{222} &= \Delta^{-}\,, \\
    T^{113} &= \frac{1}{\sqrt{3}}\Sigma^{\star +}\,, & T^{123} &= \frac{1}{\sqrt{6}} \Sigma^{\star 0}\,, & T^{223} &= \Sigma^{\star -}\,, \\
    T^{133} &= \frac{1}{\sqrt{3}}\Xi^{\star 0}\,, & T^{233} &= \frac{1}{\sqrt{3}}\Xi^{\star -}\,,\\
    T^{333} & = \Om^- \, .
\end{aligned}
\end{equation}
$T$ is again symmetric in its flavor indices and, independently, in its spin indices. 
$B$ is the matrix of baryon fields,
\begin{align}
    B=\begin{pmatrix}
        \frac{1}{\sqrt{2}}\Sigma^0 + \frac{1}{\sqrt{6}}\Lambda & \Sigma^+ & p \\
        \Sigma^- & - \frac{1}{\sqrt{2}}\Sigma^0 + \frac{1}{\sqrt{6}}\Lambda & n \\
        \Xi^- & \Xi^0 & -\sqrt{\frac{2}{3}} \Lambda
    \end{pmatrix} \, .
\end{align}

\subsection{Application in two-flavor case without \DD intermediate states \cite{Kaplan:1995yg}}
An application of the SU($2F$) symmetry was discussed in Ref.~\cite{Kaplan:1995yg}: in two-flavor QCD without any additional assumed symmetries there are 18 independent LECs associated with the 18 four-baryon dimension-6 operators describing the \Swave interactions among nucleons and {\D}'s.
From the Lagrangian in Eq.~\eqref{eq:SU4Lag}, we see that in the large-\Nc limit these 18 LECs can be expressed in terms of only $\at$ and $\bt$, i.e., there are strong constraints on the LECs in the SU(4) symmetry limit. As one example of such a constraint, Ref.~\cite{Kaplan:1995yg} showed that the \NN LECs in the $\oneS$ and $\threeS$ channels are identical in this limit:
\begin{align}
\label{eq:CNNequal}
    \CNNsing = \CNNtrip \,  
\end{align}
in the \Swave \NN Lagrangian 
\begin{align}\label{eq:NNLag}
    \calL_{\NN} = - \CNNsing (N^T P^{(\oneS)}_A N)^\dagger (N^T P^{(\oneS)}_A N)
    - \CNNtrip (N^T P^{(\threeS)}_K N)^\dagger (N^T P^{(\threeS)}_K N) \, ,
\end{align}
where the partial wave projection operators are 
\begin{align}\label{eq:NNProjectors1S0}
    P^{(\oneS)}_A & = \frac{1}{\sqrt{8}}\sigma_2 \, \tau_2\tau_A \, ,\\ 
    \label{eq:NNProjectors3S1}
    P^{(\threeS)}_K & = \frac{1}{\sqrt{8}}\sigma_2 \sigma_K \, \tau_2 \, ,
\end{align}
with $\tau_A$ ($\sigma_K$) Pauli matrices in isospin (spin) space and $A,K = 1,2,3$, using the spectroscopic notation in the superscripts: $^{2\mathscr{S}+1}L_J$, where $\mathscr{S}$ is the total spin, $L=S, P, \ldots$ is the partial wave, and $J$ is the total angular momentum.

\section{\DD intermediate states in \NN scattering}
\label{sec:DDIntermediate}

In this section we discuss the second technique upon which our results are based: the \eftnopi formalism for including intermediate \DD states in \NN scattering.  We begin with a review of the results of Ref.~\cite{Savage:1996tb} in a  different notation, present additional amplitudes, and then extend the discussion to the \threeS channel.

\subsection{\DD intermediate states in the $\oneS$ channel}
\label{sec:DDIntOneS}

Reference \cite{Savage:1996tb} analyzed \NN scattering in the \oneS channel including both nucleon and \D fields.
We summarize the parts relevant for our analysis here, but use different conventions and notation.
We write the  Lagrangian of Ref.~\cite{Savage:1996tb} in the $\oneS, I=1$ channel at LO in the \eftnopi power counting as 
\begin{align}
\label{eq:1S0Lag}
    \calL^{(\oneS)} &= \calL_{\NN}^{(\oneS)}+\calL_{\D N}^{(\oneS)}+\calL_{\DD}^{(\mathscr{S}=0,I=1)} \, ,
\end{align}
where $\calL_{B^\prime B}$ includes operators that describe $BB \leftrightarrow B^\prime B^\prime$ interactions and their corresponding LECs $C_{B^\prime B}$.\footnote{Our sign convention for $\CNDsing$ and $\CDDoneSIone$ is opposite that of Ref.~\cite{Savage:1996tb}.} 
We will only consider \Swave interactions in the \DD sector. 
Instead of labeling the \DD Lagrangian, LECs, scattering lengths, and amplitudes using superscripts in the spectroscopic notation, we will use the ($\mathscr{S}$,$I$) notation, where $\mathscr{S}$ is the total spin. For example, the $\oneS,I=1$ \DD channel in Eq.~\eqref{eq:1S0Lag} is labeled (0,1), and the $^7S_3,I=0$ \DD channel will be labeled $(3,1)$.
The \NN part of the Lagrangian is given in Eq.~\eqref{eq:NNLag}. 
The $N\D$ terms (dictating $NN \leftrightarrow \D \D$ interactions) are 
\begin{align} \label{eq:NDLagSing}
    \calL_{\D N}^{(\oneS)} = -\CNDsing (N^T P^{(\oneS)}_A N)^\dagger (\Delta^T \mathscr{P}_A^{(0,1)} \Delta) +\text{H.c.},
\end{align}
and the $\DD$ terms are 
\begin{align} \label{eq:DDLagSing}
    \calL^{(0,1)}_{\DD} = - \CDDoneSIone(\D^T \mathscr{P}_A^{(0,1)} \D)^\dagger (\D^T \mathscr{P}_A^{(0,1)} \D)\, .
\end{align}
The \NN projection operator is given in Eq.~\eqref{eq:NNProjectors1S0}. $\mathscr{P}_A^{(0,1)}$ is the corresponding projection operator for \D fields, given in App.~\ref{sec:AppLagrangians}.

The large scattering lengths in the \NN sector make it necessary to resum an infinite set of diagrams at LO in the \eftnopi power counting \cite{Kaplan:1996xu,Kaplan:1998tg, Kaplan:1998we, vanKolck:1998bw}.
While experimental information on $\NN\leftrightarrow\DD$ and \DD scattering is limited, given the close connection between $N$'s and \D's in the large-\Nc limit we also treat these interactions as nonperturbative in the following.
The infinite set of LO \Swave \NN scattering diagrams with intermediate \DD states can be calculated analytically.
Their contribution can be encoded in an effective \NN coefficient $\widetilde{C}_{\NN}^{(\oneS)}(p)$ 
\cite{Savage:1996tb},\footnote{In the following we use a modified notation from Ref.{\cite{Savage:1996tb}}, replacing $C_E^{*}$ with $\CeffSing$ and $G_E^{BB}(\vec 0,\vec 0)$ with $I_{B}(p)$, where $B=N$ or \D, and include the partial wave labels on the LECs.}
\begin{align}
\label{eq:CEsing}
    \CeffSing (p) = \CNNsing + \frac{\left[\CNDsing \right]^2 I_{\D}(p)}{1-\CDDsing I_{\D}(p)} \, ,
\end{align}
so that the leading-order (LO) amplitude including intermediate \DD states takes the form 
\begin{align}
\label{eq:Asing}
    \calA_\text{\NN,LO}^{(\oneS)} = - \frac{\CeffSing(p)}{1-\CeffSing(p) I_{N}(p)}\, .
\end{align}
$I_{N}$ ($I_{\D}$) is the integral corresponding to an \NN (\DD) loop, see Eq.~\eqref{eq:IBB} below for expressions.

While the amplitude of Eq.~\eqref{eq:Asing} has the same form as in the \D-less theory, the effective coefficient $\CeffSing(p)$ does \emph{not} correspond to a LO LEC in either the \D-less or \D-full theory. $\CeffSing(p)$ depends on the energy of the \NN system through the loop integrals and is therefore not an LEC in the usual sense. In the following we will occasionally omit the explicit $p$ dependence, distinguishing the effective coefficient by using a tilde: \CeffSing.

The Lagrangian of Eq.~\eqref{eq:1S0Lag} can also be used to calculate $\NN \leftrightarrow \DD$ and $\DD\to\DD$ scattering in the $\oneS, I=1$ channel. The calculation of \DD scattering proceeds in analogy with \NN scattering, with the role of $N$'s and \D's interchanged. The LO amplitude is 
    \begin{align}\label{A01Del}
        \calA^{(0,1)}_{\DD, \text{LO}} & = - \frac{ \CDDeffSing(p) }{ 1 - \CDDeffSing(p) I_{\D}(p)} \, ,
    \end{align}
where
    \begin{align}
     \label{CtripDel}\CDDeffSing(p) & = \CDDsing + \frac{ \left[ C_{\Delta N}^{(\oneS)}\right]^2 I_{N}(p)  }{1 - C_{\NN}^{(\oneS)} I_{N}(p)} \, .
    \end{align}
We will not be needing the $N\D \to N\D$ amplitude in what follows, but will use the $\NN \to \DD$ and $\DD \to \NN$ amplitudes, labeled $\calA_{N\D}$ and $\calA_{\D N}$, respectively, 
    \begin{align}
    \label{eq:ANDsing}
        \calA^{(\oneS)}_{N\D, \text{LO}} & = - \frac{C^{(\oneS)}_{\Delta N}}{ \left[ 1 - C_{\DD}^{(\oneS)} I_{\D}(p) \right] \left[ 1 - \CeffSing(p) I_{N}(p) \right] } \, , \\
    \label{eq:ADNsing}
        \calA^{(\oneS)}_{\D N, \text{LO}} & = - \frac{C^{(\oneS)}_{\Delta N}}{ \left[ 1 - C_{\NN}^{(\oneS)} I_{N}(p) \right] \left[ 1 - \CDDeffSing(p) I_{\D}(p) \right] } \, .
    \end{align}

Reference \cite{Savage:1996tb} discussed a few specific choices for the relative sizes of \CNNsing, \CNDsing, and \CDDsing to explore their impact on the effective \CeffSing. 
In particular, Ref.~\cite{Savage:1996tb} showed that {\D}'s decouple as the \D{}--$N$ mass splitting becomes large. 
However, as discussed above, we need to consider the opposite limit since the \D{}--$N$ mass splitting vanishes as $\Nc \to \infty$. The analysis of Ref.~\cite{Savage:1996tb} by itself therefore does not give any hints for why the large-\Nc expectations for two-nucleon processes are satisfied in an \eftnopi without \D degrees of freedom.
In the following, we extend the discussion to also include the $\threeS$ channel and investigate the relationships between the LECs in the large-\Nc limit.

\subsection{\DD intermediate states in the \threeS channel}
\label{sec:DDIntThreeS}

Because the large-\Nc analysis in the theory without explicit \D degrees of freedom relates the \oneS and the \threeS channels, we explore how robust that relationship is to inclusion of the \D by extending the analysis of $\DD$ intermediate states from the $\oneS$ channel \cite{Savage:1996tb} to the $\threeS$ channel.
The \threeS \NN, \ND, and \DD Lagrangians can be obtained from those in the \oneS channels of Ref.~\cite{Savage:1996tb} by interchanging the spin and isospin operators. 
The inclusion of intermediate \DD states in \NN scattering in the \threeS channel follows in analogy to the \oneS case.\footnote{As in the \oneS case, we restrict the discussion to LO in the EFT power counting. At higher orders we would also need to include \SD mixing.} 
The expressions for the scattering amplitudes and effective coefficients containing $\D$'s for the \threeS channels are obtained by replacing the $(\oneS)$ and $(0,1)$ labels in Eqs.~\eqref{eq:CEsing}-\eqref{eq:ADNsing} by $(\threeS)$ and $(1,0)$ labels, respectively.
In particular, the scattering amplitude can be written as 
\begin{align}
\label{eq:Atrip}
    \calA_\text{\NN,LO}^{(\threeS)} = - \frac{\CeffTrip(p)}{1-\CeffTrip(p) I_{N}(p)}\, .
\end{align}
with the effective \NN coefficient $\CeffTrip (p)$ given by
\begin{align}
\label{eq:CEtrip}
    \CeffTrip(p) = \CNNtrip + \frac{\left[ \CNDtrip \right]^2 I_{\D}(p)}{1-\CDDtrip I_{\D}(p)} \, . 
\end{align}

The LO $\threeS,I=0$ scattering amplitudes with \DD initial and/or final states are given by
    \begin{align}
        \calA^{(1,0)}_{\DD, \text{LO}} & = - \frac{ \CDDeffTrip(p) }{ 1 - \CDDeffTrip(p) I_{\D}(p)} \, ,
    \end{align}
where 
    \begin{align}
        \CDDeffTrip(p) & = C_{\DD}^{(\threeS)} + \frac{ \left[ C_{\Delta N}^{(\threeS)} \right]^2 I_{N}(p)  }{1 - C_{\NN}^{(\threeS)} I_{N}(p)} \, .
    \end{align}
The $\NN \to \DD$ and $\DD \to \NN$ amplitudes are 
    \begin{align}
        \label{eq:ANDtrip}
        \calA^{(\threeS)}_{N\D, \text{LO}} & = - \frac{C^{(\threeS)}_{\Delta N}}{ \left[1 - C_{\DD}^{(\threeS)} I_{\D}(p) \right] \left[ 1 - \CeffTrip(p) I_{N}(p) \right]} \, , \\
        \label{eq:ADNtrip}
        \calA^{(\threeS)}_{\D N, \text{LO}} & = - \frac{C^{(\threeS)}_{\Delta N}}{ \left[ 1 - C_{\NN}^{(\threeS)} I_{N}(p) \right] \left[ 1 - \CDDeffTrip(p) I_{\D}(p) \right]} \, .
    \end{align}

\section{Spin-isospin exchange symmetry in \Swave interactions}
\label{sec:flip}

Reference~\cite{Kaplan:1995yg} obtained the equality of the \NN LECs in the \oneS and \threeS channel in the large-\Nc limit by imposing an SU($2F$) symmetry on \eftnopi without explicit \D degrees of freedom. In the following we analyze the consequences of this SU($2F$) symmetry for \eftnopi including \D's.
As noted in Ref.~\cite{Kaplan:1995yg}, in the SU(4)-symmetric limit all $\NN$, \ND, and $\DD$ LECs are determined in terms of just two parameters.
These relationships can be used to compare \NN scattering in the two \Swave channels in the theory with \DD intermediate states.  
To pursue this we will need not only the expressions for spin-1/2 particles provided in Ref.~\cite{Kaplan:1995yg}, but also the expressions for \ND and \DD LECs in terms of the two SU(4) parameters. 
We will return to this framework in Sec.~\ref{sec:BBscattering}. However, there is a more direct approach for analyzing \Swave scattering.

In addition to SU(4), the Lagrangian of Eq.~\ref{eq:SU4Lag} is invariant under an exchange symmetry of spin and isospin degrees of freedom (or flip symmetry for short).\footnote{This use of the flip symmetry was proposed by M.~Wise in 2016, R.P.~Springer gave a talk on the applications to few nucleon systems at the May 9, 2017  Michigan State NSCL seminar on ``Large N constraints on pionless EFT: applications to few bodies;
a window into QCD" and on Nov.~14, 2019 at  the University of Maryland.  Further applications were discussed in Ref.~\cite{Lee_2021}.}
Application of the flip symmetry operation $F$ (with $F^\dagger = F^{-1}=F$) to Eq.~\ref{symm.field} changes the lowercase Latin superscripts (isospin labels) to Greek superscripts and the Greek subscripts (spin labels) to lowercase Latin subscripts, yielding
\begin{equation}
F \Psi^{ abc }_{\alpha \beta \gamma}F=\Psi^{ \alpha \beta \gamma }_{abc} \ \ .
\end{equation}
In particular,
\begin{equation}
\label{bil}
 F \Delta^{abc}_{\alpha \beta \gamma}  F=\Delta_{abc}^{\alpha \beta \gamma} \ \ , \ \ F N^a_\alpha F  =N^\alpha_a \ \ , \ \  {\rm and} \ \ F \epsilon^{ab} \epsilon_{\alpha \beta} F = \epsilon^{\alpha \beta} \epsilon_{ab} \, .
\end{equation}
The $F$ operation has determinant $-1$ and so is \emph{not} in the SU(4) spin-flavor group.
To obtain an element of SU(4) from $F$ requires an additional U(1) operation. 
The operators (using a labeling scheme that echoes that of their corresponding LECs) in the \Swave Lagrangians transform under $F$ as 
\begin{align}
    \ONNsing \overset{F}{\longleftrightarrow} \ONNtrip \, , \quad \ONDsing \overset{F}{\longleftrightarrow} \ONDtrip\, , \quad  \calO_{\DD}^{(0,1)} \overset{F}{\longleftrightarrow} \calO_{\DD}^{(1,0)}\, .
\end{align}
Imposing the $F$ symmetry of Eq.~\eqref{eq:SU4Lag} on Eq.~\eqref{eq:1S0Lag} and its \threeS partner, we see that not only is Eq.~\eqref{eq:CNNequal} recaptured, but we have the additional equalities in the \oneS and \threeS channels: 
\begin{align}
\label{eq:CNDequal}    
    \CNDsing & = \CNDtrip  \, , \\
\label{eq:CDDequal}    
    \CDDsing & = \CDDtrip \, .
\end{align}
(As we will see below, this flip symmetry holds even for higher spin and isospin \DD channels.)
The equality of the \NN LECs in Eq.~\eqref{eq:CNNequal}  was already derived in Ref.~\cite{Kaplan:1995yg}.
Because the LECs ${C}_{B B^\prime}$ in the two \Swave channels are equal, the effective parameters $\widetilde{C}_{\NN}^{(\oneS)}$ and $\widetilde{C}_{\NN}^{(\threeS)}$ are also equal, and so are the expressions for the scattering amplitudes, so long as the same calculational scheme is used; 
\begin{align}
\label{eq:NNAmpEqual}
    \calA_{\NN,\text{LO}}^{(\oneS)} = \calA_{\NN,\text{LO}}^{(\threeS)} \, .
\end{align}
Recall that these are the amplitudes obtained for the theory \emph{with} explicit \D degrees of freedom.
In the theory \emph{without} {\D}'s, the expression for the amplitude in the $s$ channel is given by \cite{Kaplan:1998we}
\begin{align}
\calA_{\text{LO},\slashed{\Delta}}^{(s)} = -\frac{C_{\NN,\slashed{\D}}^{(s)}}{1 - C_{\NN,\slashed{\D}}^{(s)} I_N} \, ,
\end{align}
where $C_{\NN,\slashed{\D}}^{(s)}$ is the LO \NN LEC in the \D-less theory. 
Since the amplitudes are related to observables, we can require that for a given channel the LO amplitudes in the theories with and without intermediate \D's are identical, i.e.,
\begin{align}
    \calA_{\NN,\text{LO}}^{(s)} = \calA_{\text{LO},\slashed{\Delta}}^{(s)} \, .
\end{align}
The equality of the \oneS and \threeS amplitudes of Eq.~\eqref{eq:NNAmpEqual} in the large-\Nc limit means that the amplitudes $\calA_{\text{LO},\slashed{\Delta}}^{(s)}$ in the \D-less theory are also identical, which in turn implies that the \oneS and \threeS LECs in the \D-less theory have to be equal to each other, 
\begin{align}
\label{eq:CDless}
    \Czerosing = \Czerotrip \, .
\end{align}
This is the same result as obtained by applying the large-\Nc scaling rules directly to the \D-less theory.
This does not mean that the values of the LECs in the \D-full  theory are the same as those in the \D-less theory; what it shows is that in each of these theories the spin-singlet and spin-triplet LECs remain equal to each other in the large-\Nc limit.

This analysis also holds for \Swave interactions at the next  order in the \eftnopi power counting. 
The corresponding Lagrangian contains two-derivative operators, but the indices of the derivatives are contracted with each other and \emph{not} with the indices of spin operators; these derivative operators are unaffected by the application of the flip symmetry operation, leading to the equality of \oneS and \threeS LECs associated with two-derivative operators. 
As a result, the \NN \Swave scattering amplitudes in the two channels are again identical at this order. Requiring the amplitudes to match those in the \D-less theory, the two-derivative \Swave LECs in the \D-less theory are then also expected to be equal to each other.
This expectation is in agreement with the large-\Nc analysis in the \D-less theory of Ref.~\cite{Schindler:2018irz}. 
The equality of the \Swave LECs in the \D-full theory for $NN$, $NN \leftrightarrow \D \D$, and $\DD$ scattering follows from flip symmetry even at higher orders in the \eftnopi expansion. However, \threeS scattering amplitudes at higher order will include contributions from \SD mixing, so that the arguments used at LO and NLO to analyze the LECs in the \D-less theory will not apply. We explore \SD mixing further in Sec.~\ref{sec:SDmixing}.

\section{Higher partial waves}
\label{sec:Higher}

\subsection{$P$ waves and other perturbative \NN channels}
\label{sec:Pwaves}

Large-\Nc analyses have also been applied to \NN interactions beyond $S$ waves. 
As seen above, in \eftnopi the \D degrees of freedom contribute to  \NN interactions through loop diagrams. 
In the \D-less theory, \NN interactions for 
$L \ge 1$ are treated perturbatively; loop diagrams  do not contribute at the same order as the first tree-level diagrams in these higher partial wave channels (see, e.g., Refs.~\cite{Kaplan:1998sz,Rupak:1999rk}). 
Exceptions include interactions that mix {\Swave}s with higher partial waves, e.g., \SD mixing and parity-violating (PV) \SP interactions. This is discussed below.
It is possible that the power counting in the \D-full theory is different from that in the \D-less theory and that loop diagrams have to be resummed for some $L \ge 1$ as well if \D degrees of freedom are included dynamically. 
However, given that the theory that results from integrating out the \D is apparently perturbative, it seems likely that interactions with $L \ge 1$ are also perturbative in the \D-full theory.
This means that at sufficiently low order in the \eftnopi power counting, the inclusion of the \D does not significantly impact the \NN LECs in $L\ge 1$ waves. 
In this case, we expect that the relationships among LECs for $L\ge 1$ interactions in the \D-full theory are not dramatically changed from those within the \D-less theory.

Reference \cite{Schindler:2018irz} considered several relationships between \Swave, \Pwave, and \SD-mixing two-derivative LECs in the \D-less theory. A relationship that was not shown in Ref.~\cite{Schindler:2018irz}, but that follows from that analysis is 
\begin{align}\label{eq:TwoDerRel}
    \frac{5 \CPtwo +\frac{1}{2} \left( \CPone - \ConeP \right)}{C^{(s)}_{2,\slashed{\D}}}= 1 ,
\end{align}
which holds for $s=\oneS$ and \threeS with expected corrections of $O(1/\Nc)$, given the expected equality of the \Swave LECs at leading order. 
This relationship is derived based on a large-\Nc analysis in the \D-less theory. As discussed above, the \Pwave LECs are not expected to be impacted significantly by the inclusion of the \D, so the notation for the \Pwave LECs is not modified in the equation above. 
However, the \Swave \NN LECs in the theories with and without the \D can be very different. If the \D had a large impact on the \NN LECs, it is not clear why relationships like that in Eq.~\eqref{eq:TwoDerRel} should hold. Using the values for the LECs given in Ref.~\cite{Schindler:2018irz} yields 
\begin{equation}\label{eq:PwaveRatios}
\begin{split}
    \frac{5 \CPtwo +\frac{1}{2} \left( \CPone - \ConeP \right)}{C^{(\threeS)}_{2,\slashed{\D}}(\mu = m_\pi)} &\approx 1.1 ,\\
    \frac{5 \CPtwo +\frac{1}{2} \left( \CPone - \ConeP \right)}{C^{(\oneS)}_{2,\slashed{\D}}(\mu = m_\pi)} &\approx 1.7\, .
\end{split}
\end{equation}
These ratios exhibit strong $\mu$ dependence, reflecting the well-known strong $\mu$ dependence of the \Swave LECs in the power divergence subtraction (PDS) scheme \cite{Kaplan:1998tg} used in Ref.~\cite{Schindler:2018irz}.
The relative closeness to the naive (\D-less) large-\Nc expectation of Eq.~\eqref{eq:TwoDerRel} may be an indication that the effects of including the \D are relatively minor even for the \Swave \NN LECs.

\subsection{\SD mixing}
\label{sec:SDmixing}
\SD mixing and PV $S$-$P$ processes, because their \Swave component is nonperturbative, are not subject to the perturbative arguments of the previous section. They will require a different analysis to understand the impact of including the \D degree of freedom.
In this subsection we discuss \SD mixing before considering PV \SP interactions in the next subsection.

\begin{figure}
    \begin{center}
\includegraphics[width=\textwidth]{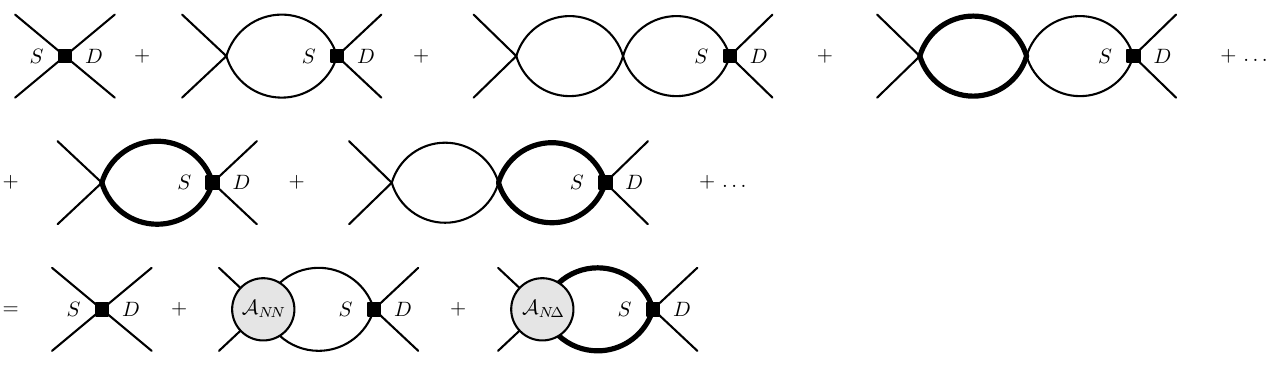}
    \end{center}
\caption{Diagrams contributing to \SD mixing. Thin lines denote nucleons and thick lines denote \D's. The squares stand for \SD mixing vertices involving either \NN's and/or \DD's as indicated by the thickness of the attached legs. The gray circles with amplitude labels inside represent LO \threeS scattering including \D intermediate states.}
    \label{fig:SD}
\end{figure}

In the \D-less theory, there is one two-derivative operator that induces \SD mixing, multiplied by an LEC \CsdDless. In addition to the tree-level diagram with this vertex, diagrams with additional LO \threeS interactions appear at the same order in the power counting and have to be resummed.
The \D-full theory contains an analogous \NN \SD operator $\calO_{\NN}^{\SD}$, with the associated LEC \CsdNN. 
Again, diagrams with additional \Swave interactions have to be resummed, and \DD intermediate states contribute to the $S$-wave rescattering, as discussed in Sec.~\ref{sec:DDIntThreeS} and shown in Fig.~\ref{fig:SD}. 
However, there can also be an operator $\calO_{\D N} ^{\SD}$ with LEC \CsdNDelta that connects \Swave \DD states to \Dwave \NN states.\footnote{There can also be \SD operators that connect \Swave \NN and \DD states to \Dwave \DD states. However, when considering \NN initial and final states, diagrams containing these operators require an additional insertion of a \Dwave operator to return to a final \NN state and are therefore of higher order.}
There is therefore a second class of diagrams that contribute to \SD transitions for diagrams with \NN initial and final states, in which the incoming \Swave \NN state is scattered into an \Swave \DD state that then connects to the \Dwave \NN state through the $\calO_{\D N}^{\SD}$ operator.
The \Swave part of the diagrams can be  resummed and collected in the LO \Swave amplitudes for $\NN$ and $\NN \to \DD$ scattering. 
Using the expressions of Eqs.~\eqref{eq:Atrip} and \eqref{eq:ANDtrip} for the \Swave amplitudes, the amplitude for \SD mixing is proportional to 
\begin{align}
    \calA^{(\SD)}_{\NN,\text{LO}} &\propto\frac{1}{1-\CeffTrip I_N} \left[ \CsdNN + \CsdNDelta \frac{\CNDtrip I_{\D}}{1-\CDDtrip I_{\D}} \right] = \frac{\Ceffsd}{\CeffTrip} \calA_{\NN,\text{LO}}^{(\threeS)}\, ,
\end{align}
where an effective coefficient is introduced, just as it was in the \Swave case,
\begin{align}
\label{eq:CSDeff}
    \Ceffsd(p) \equiv \left[ \CsdNN + \CsdNDelta \frac{\CNDtrip I_{\D}}{1-\CDDtrip I_{\D}} \right] \, .
\end{align}
The calculation in the \D-less theory gives 
\begin{align}
    \calA^{(\SD)}_{\text{LO},\slashed{\D}} 
    &\propto \frac{\CsdDless}{\Czerotrip} \calA_{\text{LO},\slashed{\D}}^{(\threeS)} \, .
\end{align}
Requiring that the LO \SD amplitudes match and using the equality of the LO \Swave amplitudes in the \D-full and \D-less theories yields
\begin{align}
\label{eq:SDratio}
    \frac{\Ceffsd}{\CeffTrip} = \frac{\CsdDless}{\Czerotrip} \, .
\end{align} 
The effective coefficients \Ceffsd and \CeffTrip are not LECs themselves, but are functions of the LECs in the \D-full theory.
Therefore, Eq.~\eqref{eq:SDratio} does \emph{not} indicate that the ratio of \SD to \threeS LECs in the \D-less theory is identical to that in the \D-full theory (for which the large-\Nc analysis is expected to hold) unless \CNDtrip is small compared to the other LECs. That is, as $\CNDtrip \to 0$, the effective coefficients \Ceffsd and \CeffTrip reduce to the LECs \CsdNN and \CNNtrip, respectively.

While the large-\Nc relationships among LECs obtained in \D-less \eftnopi are in general agreement with data for $S$ and $P$ waves, this is evident only if the renormalization point is chosen so that the relationships are not obscured by the finite scattering lengths in the $S$-wave \NN channels (see discussions in Sec.~\ref{sec:Pwaves} and Ref.~\cite{Schindler:2018irz}).  
One way that the result of this section, Eq.~\eqref{eq:SDratio}, differs from those examples, is that Eq.~\eqref{eq:SDratio} is $\mu$-independent.  
Even though the physics in this channel involves unnaturally large scattering lengths, that dependence cancels in the ratio of  Eq.~\eqref{eq:SDratio}.  
This therefore may be considered a more robust result. However, Ref.~\cite{Schindler:2018irz} found that inclusion of the \SD mixing term proportional to \CsdDless 
challenges the experimental agreement with the ordering of LECs suggested by the large-\Nc analysis.\footnote{Reference~\cite{Schindler:2018irz} labels this LEC $C^{(SD)}$.}
This was traced to the fact that the value of \CsdDless extracted from experiment is unnaturally small compared to the expectation from power counting in \eftnopi \cite{Chen:1999vd}. 
It is possible that the inclusion of explicit \D degrees of freedom might address this potential discrepancy with the naive large-\Nc expectation as applied in the \D-less theory.
However, as seen in the previous subsection and as discussed further in Sec.~\ref{sec:BBscattering}, there are some indications that the $C_{\D N}$ are small.
In that case, Eq.~\eqref{eq:CSDeff} shows that the mixing term in the \D-full theory approaches that in the \D-less theory.

That the \SD-mixing term might not conform to the large-\Nc expectation does not mean that the large-\Nc analysis is not useful in general. 
Given that the expansion parameter can be as large as 1/3, factors  that appear for reasons not captured by the SU(4) symmetry can lead to a re-ordering of operator contributions outside of large-\Nc expectations.
This is why it is useful to think of large-\Nc ordering as indicating trends more than rigid predictions.

\subsection{Parity-violating \SP interactions}
\label{sec:parity}

The experimental situation is not as clear for PV observables as for parity-conserving (PC) observables because the signals are small and the experiments are difficult.  
Some authors \cite{Haxton:cipanp18,Vanasse:2018buq} have argued that the available constraints point towards inconsistencies with the large-\Nc expectations of Ref.~\cite{Schindler:2015nga}. 
The analysis of Ref.~\cite{Haxton:cipanp18} combines few-body observables with those in heavier nuclear systems, such as $^{18}\text{F}$. Using the LECs from \eftnopi in nuclear calculations introduces uncontrolled theoretical uncertainties. In particular, it is inconsistent to use the LECs extracted using one EFT and use them in a different theory. Reference \cite{Vanasse:2018buq} avoids these complications by restricting the discussion to two- and three-nucleon observables. However, the experimental results used in this analysis have large uncertainties, which translate in large uncertainties on the extracted values of the LECs.
At present, what we can say is that available data on both two- and three-nucleon systems are not inconsistent with the large-\Nc constraints imposed on the \D-less version of \eftnopi.

\begin{figure}
    \begin{center}
\includegraphics[width=0.8\textwidth]{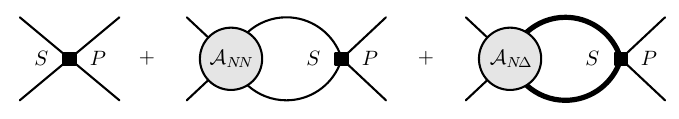}
    \end{center}
    \caption{Diagrams contributing to parity-violating \SP mixing in the theory with \D degrees of freedom. Thin lines denote nucleons and thick lines denote Deltas. The square stands for an \SP mixing vertex. The gray circles represent the (parity-conserving) \Swave scattering amplitudes including \D intermediate states; in particular the LO \Swave amplitudes $\calA_{N\!N} \ne \calA_{N\!N,\slashed{\Delta}}$.
    \label{fig:PV}}
\end{figure}

PV interactions that mix $S$ and $P$ waves provide another example in which nonperturbative \Swave interactions have to be resummed. 
The calculation that includes the \D intermediate states is analogous to that for \SD mixing. All LO \SP PV processes can be expressed in terms of the PC \Swave scattering amplitudes for $\NN \to \NN$ and $\NN \to \DD$ and the PV LECs \CPVNN and \CPVNDelta, where \SP denotes any of the five allowed transitions; see Fig.~\ref{fig:PV}. 
Using the expressions for the amplitudes $\calA_{\NN,\text{LO}}^{(S)}$ and $\calA_{\ND,\text{LO}}^{(S)}$ given in Eqs.~\eqref{eq:Asing}/\eqref{eq:Atrip} and \eqref{eq:ANDsing}/\eqref{eq:ANDtrip}, respectively, the \SP amplitude in the theories with explicit \D's are proportional to
\begin{align}
    \calA^{(\SP)} &\propto \frac{1}{1-\Ceff(p) I_N} \left[ \CPVNN + \CPVNDelta\frac{ C^{(S)}_{\Delta N} I_{\D}}{1-C^{(S)}_{\DD} I_N} \right] \\
    &= - \frac{\CPVeff(p)}{\Ceff(p)}  \calA_\text{\NN,LO}^{(S)}\, .
\end{align}
where, analogous to the PC case, we introduce an effective PV coefficient
\begin{align}
\label{eq:CeffPV}
    \CPVeff(p) \equiv \left[ \CPVNN + \CPVNDelta\frac{ C^{(S)}_{\Delta N} I_{\D}}{1-C^{(S)}_{\DD} I_N} \right] \, .
\end{align}

The corresponding amplitude in the \D-less theory can be calculated from the first two diagrams in Fig.~\ref{fig:PV} and is given by 
\begin{align}
    \calA^{(\SP)}_\slashed{\D} &= \frac{1}{1-C^{(S)}_{\NN,\slashed{\D}} I_N} \CPVDless \\
    &= -\frac{\CPVDless}{C^{(S)}_{\NN,\slashed{\D}}} \calA_{\text{LO},\slashed{\D}}^{(S)}\, .
\end{align}
To compare to experiment we  require that the PV amplitudes in the theories with and without \D's are equal. Using that the PC amplitudes also match yields
\begin{align}\label{eq:PV_DfullDless}
    \frac{\CPVeff(p)}{\Ceff(p)} = \frac{\CPVDless}{C^{(S)}_{\NN,\slashed{\D}}} \, .
\end{align}
Comparing the relative sizes of two PV LECs in the \D-less theory corresponds to considering the ratio of the two LECs. Because of the equality of the two \Swave LECs $C^{(S')}_{\NN,\slashed{\D}} = C^{(S)}_{\NN,\slashed{\D}}$ this ratio can be written as
\begin{align}
    \frac{\CPVDlessprime}{\CPVDless} = \frac{\CPVDlessprime}{C^{(S')}_{\NN,\slashed{\D}}}\frac{C^{(S)}_{\NN,\slashed{\D}}}{\CPVDless} \, .
\end{align}
Using Eq.~\eqref{eq:PV_DfullDless} and the equality of the effective parameters in the two \Swave channels gives
\begin{align}
\label{eq:pvresult}
    \frac{\CPVDlessprime}{\CPVDless} &= \frac{\CPVDlessprime}{C^{(S')}_{\NN,\slashed{\D}}}\frac{C^{(S)}_{\NN,\slashed{\D}}}{\CPVDless} \nonumber \\
    &=\frac{\CPVeffprime(p)}{\Ceff(p)} \frac{\Ceffprime(p)}{\CPVeffprime(p)} \nonumber \\
    &=\frac{\CPVeffprime(p)}{\CPVeff(p)} \, .
\end{align}
Once again this result is $\mu$-independent, a ratio that reflects the large-\Nc ordering that does not rely on carefully chosen subtraction points.
We also see from Eq.~\eqref{eq:CeffPV} that the effective parameters \CPVeff reduce to the \NN LECs in the theory with \D's as $\CDN^{(S)}$ vanishes.
That the large-\Nc constraints imposed on the \D-less theory are not in conflict with data might be an indication that including the \D as an explicit degree of freedom does not impact the values of the LECs dramatically, similarly to the discussion below Eq.~\eqref{eq:PwaveRatios}. We will discuss other indications that $\CDN^{(S)}$ might be small in Sec.~\ref{sec:BBscattering}.

\section{SU(2$F$) constraints on the \D-full pionless Lagrangian}
\label{sec:BBscattering}

As noted in Ref.~\cite{Kaplan:1995yg}, in the SU(4)-symmetric limit all $\NN$, \ND, and $\DD$ LECs are determined in terms of two parameters, which we label $\at$ and $\bt$ in our convention; see Eq.~\eqref{eq:SU4Lag}. 
The authors of Ref.~\cite{Kaplan:1995yg} focused on spin-1/2 particles and provided expressions for the $NN$ LECs, which in terms of \at and \bt are
\cite{Kaplan:1995yg}:
\begin{align}
\label{eq:CNNab}
    \CNNsing = 2 \left( \at - \frac{\bt}{27} \right) = \CNNtrip \, .
\end{align}
Because the singlet and triplet $NN$ LECs are equal in the SU(4) limit, information from \NN scattering is not sufficient to determine both \at and \bt.
We therefore also consider $\NN\leftrightarrow \DD$ and \DD scattering in all allowed \Swave channels.
$\calL_{\D N}^{(\oneS)}$ is given in Eq.~\eqref{eq:NDLagSing}, and the corresponding \threeS Lagrangian can be obtained by interchanging spin and isospin operators as discussed in Sec.~\ref{sec:DDIntThreeS}, yielding\footnote{We do not consider $N\D\to N\D$ scattering since that is not needed for this analysis.}
\begin{align} \label{eq:NDLagShort}
    \calL_{\D N} = -\CNDsing (N^T P^{(\oneS)}_A N)^\dagger (\Delta^T \mathscr{P}_A^{(0,1)} \Delta) 
    -\CNDtrip (N^T P^{(\threeS)}_K N)^\dagger (\Delta^T \mathscr{P}_{K}^{(1,0)} \Delta)+\text{H.c.} \, .
\end{align}
The \DD Lagrangian is the generalization of Eq.~\eqref{eq:DDLagSing} 
\begin{align} \label{eq:DDLagShort}
    \calL_{\DD} = -\sum_{\mathscr{S},I} C_{\DD}^{(\mathscr{S},I)}(\D^T \mathscr{P}^{(\mathscr{S},I)} \D)^\dagger (\D^T \mathscr{P}^{(\mathscr{S},I)} \D)\, 
\end{align}
where the sum runs over all Pauli-allowed combinations of spin $\mathscr{S}$ and isospin $I$.
The \NN projection operators $P^{(\oneS)}_A$ and $P^{(\threeS)}_K$ are given in Eqs.~\eqref{eq:NNProjectors1S0} and \eqref{eq:NNProjectors3S1}, while the $\mathscr{P}^{(\mathscr{S},I)}$ are operators that project onto \DD $S$-wave channels with total spin $\mathscr{S}$ and isospin $I$, given in App.~\ref{sec:AppLagrangians}.
The $C_{\DD}^{(\mathscr{S},I)}$ are the LECs for each channel.

The $\D N$ and $\DD$ LECs can also be expressed in terms of \at and \bt.
The expressions for the $\D N$ partial-wave LECs are 
\begin{align}
\label{eq:CNDab}
    \CNDsing = \frac{8\sqrt{5}}{27}\bt = \CNDtrip\, , 
\end{align}
while those for the \DD LECs are 
    \begin{alignat}{2}
    \label{eq:CDDabNNchannel:1}
        \CDDoneSIone &= 2\left(\at + \frac{\bt}{27}\right) &= \CDDthreeSIzero \, , \\
    \label{eq:CDDabNNchannel:2}
        \CDDoneSIthree & = 2\left(\at - \frac{\bt}{3}\right)  &= \CDDsevenSIzero \, , \\
    \label{eq:CDDabNNchannel:3}    
        \CDDthreeSItwo & = 2\left(\at - \frac{\bt}{27}\right)  &= \CDDfiveSIone \, ,\\
    \label{eq:CDDabNNchannel:4}
        \CDDfiveSIthree & = 2\left(\at + \frac{\bt}{3}\right) &= \CDDsevenSItwo \, .
\end{alignat}
These results show the symmetry of the baryon-baryon interactions under the interchange of spin and isospin discussed in Sec.~\ref{sec:flip}.

Note that if \bt=0, all of the above LEC's remain non-zero except for \CDN; \NN and \DD scattering can remain nonperturbative but the \NN and \DD sectors decouple.
Further,  the \bt contribution to the \NN and \DD LECs  involved in \NN scattering (see Eqs.~\eqref{eq:CNNab} and \eqref{eq:CDDabNNchannel:1}) is suppressed by a factor of 1/27 relative to \at. If \bt is small, or if its impact is small, this would explain why the large-\Nc constraints naively applied to \D-less \eftnopi are supported by experiment.

In addition to decoupling the \NN and \DD sectors, the $\bt = 0$ limit has a non-trivial effect on the SU(4) symmetry exhibited by the \NN and \DD system. 
Considering the \NN sector alone, the two \NN LECs are equal for any value of \bt;  setting $\bt=0$ does not lead to an enlarged symmetry. 
But an enlarged symmetry \emph{does} emerge when considering both \NN and \DD interactions; for \bt = 0, \emph{all} \NN and \DD LECs are equal to each other, corresponding to an SU(20) symmetry. 
From the large-\Nc perspective alone, \bt can take any value. 
Below we will explore indications that \bt might be close to zero when we consider the unitary limit.

\subsection{Higher spin/isospin channels in \Swave \DD scattering}

While all \NN, $\D N$, and \DD LECs are determined in terms of the SU$(2F)$ parameters \at and \bt, the symmetry group does not constrain the size of these parameters.  To obtain them we need experimental or lattice input beyond the \Swave \NN sector because of the equality of the \oneS and \threeS scattering amplitudes under SU$(2F)$.
It is convenient (but not necessary) to use a \DD amplitude from a channel that is not coupled to the \NN sector, such as the $S=3, I=1$ channel.
Limiting the discussion to scattering at zero momentum, we can express the SU$(2F)$ parameters $\at$ and $\bt$ in terms of two scattering lengths, which are defined through the zero-momentum limit of the scattering amplitudes with identical initial and final states, 
\begin{align}
\label{eq:ScatteringLengthDef}
        \lim_{p \to 0} \calA^{(\mathscr{S},I)}_{B B}(p) \to - \frac{4 \pi}{M_B} a^{(\mathscr{S},I)}_{B B} \, .
    \end{align}

To calculate these amplitudes in \D-full \eftnopi requires expressions for the loop integrals. They are given by 
\begin{align}
\label{eq:IBB}
    I_{B}(p) & = \int \frac{d^d q}{(2\pi)^d}\frac{1}{E-q^2/M_B + i\epsilon} = - \frac{M_B}{4\pi}(\mu + ip)\, , 
\end{align}
where $B$ denotes the baryon with mass $M_B$ and the expressions on the right are the results obtained in the power divergence subtraction (PDS) scheme \cite{Kaplan:1998tg} with $\mu$ the renormalization scale.\footnote{The definition of $I_{\D}$ differs from the corresponding integral in Ref.~\cite{Savage:1996tb}, in which the \D mass is expressed in terms of the nucleon mass $M_N$ and the \D{}--$N$ mass splitting.  In the large-\Nc limit considered here, the \D{}--$N$ mass splitting vanishes, i.e., $M_N=M_{\D}$.} 
Reference~\cite{Savage:1996tb} evaluated the \NN and \DD loop integrals using minimal subtraction (MS). 
This corresponds to setting $\mu = 0$ in Eq.~\eqref{eq:IBB}.
There are two reasons to use the PDS scheme. First, as was shown in Ref.~\cite{Kaplan:1998tg}, using the PDS scheme provides a power counting for the \NN system that justifies the resummation of the two-baryon bubble chains, as required by the anomalously large scattering lengths in the \NN system.
Second, previous large-\Nc analyses of \NN LEC's showed that the large-\Nc expectations  are not satisfied for values of $\mu\ll \mpi$ \cite{Kaplan:1995yg,Schindler:2018irz} 
The $\mu$ dependence of $I_{N}$ and $I_{\D}$ means that the SU(4) parameters are also $\mu$-dependent, i.e., $\at = \at(\mu)$ and $\bt = \bt(\mu)$ when matched to \D-full pionless EFT.

For \DD scattering in the $\mathscr{S} = 3$, $I=0$ channel, since only intermediate \DD states contribute, the amplitude is 
\begin{align}
\label{eq:DDAmp30}
    \calA_{\DD,\text{LO}}^{(3,0)} = - \frac{\CDDsevenSIzero(\mu) }{1 -\CDDsevenSIzero(\mu) I_{\D} } \, .
\end{align}
The LEC $\CDDsevenSIzero(\mu)$ can be expressed in terms of the scattering length in this channel, 
\begin{align}
   \label{eq:LECgeneral} \CDDsevenSIzero(\mu) = \frac{4\pi}{M_{\D}}\frac{1}{\frac{1}{a_{\DD}^{(3,0)}}-\mu} \, .
\end{align}
Analogous expressions hold for the LECs in other channels that do not receive contributions from additional intermediate baryon states \cite{Kaplan:1998we}. 
For channels with contributions from other baryon intermediate states, a corresponding expression holds for the effective parameters $\widetilde C_{BB}$. For example, for the singlet channel \NN scattering considered in Eq.~\eqref{eq:Asing}, we have
\begin{align}
\label{eq:CeffGeneral}
\CeffSing=\frac{4 \pi}{M_N}\frac{1}{\frac{1}{\asing} - \mu} \, .
\end{align}
A parallel expression holds in the \threeS channel with the replacement $\asing \to \atrip$.
The large-\Nc expectation that $\CeffSing = \CeffTrip$ does not imply that $\asing=\atrip$; since the LECs are not observables and are $\mu$-dependent, the equality is approximately satisfied as long as $\mu$ is appropriately chosen \cite{Kaplan:1995yg}. 
For the $S$ waves, this means $\mu\gg 1/\aNN$, where \aNN is one of the physical scattering lengths \asing or \atrip.

Combining the \NN amplitude with the \DD amplitude in the $\mathscr{S} = 3$, $I=0$ channel using the truncated notation $a_{\DD}^{(3,0)} \to a_3$ yields 
\begin{align}
    \tilde a(\mu) & = \frac{\pi }{M_\Delta }   \frac{- M_\Delta  ( \mu - \frac{1}{\aDDthreezero}) (10 \mu - \frac{9}{\aDDthreezero}) +   M_N (10\mu - \frac{1}{\aDDthreezero}) (\mu-\frac{1}{\aNN})} 
                         {   (\mu - \frac{1}{\aDDthreezero}) \left[ 5  \mu  M_\Delta  (\mu - \frac{1}{\aDDthreezero})-M_N  (5  \mu + \frac{4}{\aDDthreezero}) (\mu - \frac{1}{\aNN})\right]} \, ,\\
    \tilde b(\mu) & = \frac{27 \pi }{M_\Delta \aDDthreezero}  \frac{M_\Delta (\mu -\frac{1}{\aDDthreezero}) - M_N (\mu -\frac{1}{\aNN})}
    {  (\mu -\frac{1}{\aDDthreezero}) 
\left[5  \mu 
   M_\Delta ( \mu - \frac{1}{\aDDthreezero})-M_N (5 \mu + \frac{4}{\aDDthreezero})
   (\mu - \frac{1}{\aNN})\right]} \, .
\end{align}
With \at and \bt determined in terms of scattering lengths in two independent channels, we can insert them into the  amplitude  expressions for other channels.
These amplitudes will be $\mu$-dependent if $M_N \neq M_\Delta$, i.e., using non-degenerate masses with an SU(4) invariant Lagrangian leads to amplitudes that are not renormalization group invariant.
For example, the \DD amplitude at $p=0$ in the $\mathscr{S} = 0, I = 1$ channel becomes
    \begin{align}
        \calA_{\DD,\text{LO}}^{(0,1)} & = \pi \frac{\left(-5 \aNN M_\Delta^2 + \aDDthreezero M_\Delta M_N \right) - 5 \mu \delta_M \left( -2 \aNN M_\Delta + \aDDthreezero M_N \right) - 5 \mu^2 {\aDDthreezero}^2 \aNN \delta_M^2}{M_\Delta^2 M_N} \, ,
    \end{align}
where $\delta_M = M_{\D}-M_N$ is the \D{}--$N$ mass splitting.
The second and third terms in the above equation, which are the $\mu$-dependent terms, are both proportional to $\delta_M$ and will vanish in the degenerate limit.
This conclusion holds even for nonzero $p$.
If we kept nondegenerate baryon masses, we would also need to include SU(4)-breaking interactions.
With degenerate masses $M_N = M_{\D} =M$, the expressions for \at and \bt simplify to 
\begin{align}
\label{eq:atSol}
    \tilde a(\mu) & = - \frac{\pi }{M \aDDthreezero}   \frac{\left[ 2 \aDDthreezero \mu  (9 \aNN - 5 \aDDthreezero)-9 \aNN+\aDDthreezero\right]}{(\mu - \frac{1}{\aDDthreezero}) (9 a_{\NN} \mu -5 \aDDthreezero \mu -4)} \, ,\\
    \label{eq:btSol}
    \tilde b(\mu) & = - \frac{27 \pi }{M \aDDthreezero}   \frac{\aDDthreezero - \aNN}{(\mu - \frac{1}{\aDDthreezero}) (9 a_{\NN} \mu -5 \aDDthreezero \mu -4)} \, .
\end{align}

With $\at$ and $\bt$ fixed in terms of $a_{\NN}$ and $\aDDthreezero$, and the degenerate limit constraint imposed, all other \DD amplitudes can be expressed in terms of these two scattering lengths. For example, the \DD scattering lengths in the $\oneS, I=1$ and $\threeS,I=0$ channels are
  \begin{align}
        a_{\DD}^{(0,1)} = a_{\DD}^{(1,0)} = \frac{1}{4} \left( 5 a_{\NN} - \aDDthreezero \right) \, .
    \end{align}

Unlike the \NN scattering lengths, the \DD scattering lengths are currently not known. 
\DD interactions have received some renewed attention recently, due to the possible existence of a \DD bound state in the $J=3$, $I=0$ channel, the $d^*(2380)$, see for example Ref.~\cite{Clement:2016vnl} for a review.
The HAL QCD collaboration \cite{Gongyo:2020pyy} performed a lattice QCD calculation of the \DD potential in this channel for heavy quark masses and extracted phase shifts and binding energies for a \DD bound state. At the considered quark masses (at which the \D is stable against decay into $N\pi$) the extracted binding energies are on the order of 30 MeV, indicating a relatively deep bound state.
Currently, no lattice determination of the \DD scattering lengths is available.

\subsection{The unitary limit}
\label{sec:unitary}

References~\cite{Braaten:2003eu,Kievsky:2015zga,Kievsky:2015dtk,Konig:2016utl,Teng:2024exc}, for example, have explored the so-called unitary limit in nuclear systems and found not only important simplifications but that nature apparently lives sufficiently close to the unitary limit that it is useful to expand about that limit.  
In the unitary limit, the scattering length is infinite, the other effective range parameters vanish, and the baryon-baryon scattering amplitude is purely imaginary.
Taking the unitary limit for \NN scattering (see, e.g., Eq.~\eqref{eq:CeffGeneral}) and using Eqs.~\eqref{eq:CNNab}, \eqref{eq:CNDab}, and \eqref{eq:CDDabNNchannel:1}, we can solve for \bt as a function of \at, yielding 
\begin{align}
\label{eq:bNNunitary}
    \bt(\mu) = \pm \left( 3\at(\mu) +\frac{6\pi}{M \mu} \right) \, .
\end{align}
There are two solutions since \bt enters \Ceff quadratically through \CDN.
Inserting the positive solution into the expression for \DD scattering  in the $(3,0)$ channel results in an amplitude that is \emph{also} in the unitary limit, independent of \at. 
However, for other channels the resulting amplitude is $\mu$-dependent unless 
\begin{align}
\label{eq:atspecial}
    \at(\mu) = -\frac{2\pi}{M \mu}\, ,
\end{align}
in which case the amplitudes are again in the unitary limit,
\begin{align}
        \calA & = \frac{4 \pi i}{M p} \, .
    \end{align}
Further, for this value of $\at$, Eq.~\eqref{eq:bNNunitary} gives
\begin{align}
\label{eq:btzero}
    \bt(\mu) = 0 \, .
\end{align}
The negative solution in Eq.~\eqref{eq:bNNunitary} yields the same outcome. 
For example, the $\mathscr{S} = 3, I = 0$ \DD amplitude of Eq.~\eqref{eq:DDAmp30} is $\mu$-dependent unless Eq.~\eqref{eq:atspecial} is recaptured, which again yields
$
\bt(\mu) = 0 \, .
$
For these values of $\at$ and $\bt$, \emph{all} \DD amplitudes are in the unitary limit. Further, since $\bt=0$, Eq.~\eqref{eq:CNDab} shows that $\CNDsing=\CNDtrip = 0$, the \NN and \DD sectors decouple and the theory is invariant under an SU(20) symmetry.

We could instead begin by taking the unitary limit in the $\mathscr{S} = 3, I=0$ \DD scattering channel without specifying the \NN scattering length. (The \DD scattering lengths have not been determined from experiment or lattice QCD.)
In the limit $\aDDthreezero \to \infty$, Eqs.~\eqref{eq:atSol} and \eqref{eq:btSol} simplify to Eqs.~\eqref{eq:atspecial} and \eqref{eq:btzero}, the same result as obtained from starting in the \NN unitary limit. The \NN amplitudes and other \DD amplitudes also take their unitary limit form, the \NN and \DD sectors decouple, and the baryon interactions show an SU(20) symmetry.

We conclude that if nature is indeed perturbatively close to the unitary limit, it is not surprising that large-\Nc expectations applied to theories without an explicit \D tend to agree with experiment, because the \DD sector decouples. This gives us more confidence in the procedure when applied to physics that is currently under-constrained by measurements, such as PV and dark matter searches \cite{Richardson:2021liq}.

Equation~\eqref{eq:btSol} shows that the decoupling of the \NN and \DD sectors and the resulting invariance under SU(20) does not require the unitary limit; the parameter \bt also vanishes for finite scattering lengths if $\aDDthreezero = \aNN$.
This is  reminiscent of the finding in the \NN sector, where the emergence of Wigner symmetry follows from $\asing=\atrip$ without the scattering lengths taking their unitary limit values \cite{Mehen:1999qs}.

The impact of $\bt=0$, whether arising from the unitary limit or the equality of scattering lengths, is that $C_{\Delta N}=0$.
We can see that
for \SD mixing (Eq.~\eqref{eq:SDratio}) or PV (Eq.~\eqref{eq:pvresult}), the \D-less \eftnopi results prevail.
At the same time, these sources for $\bt=0$ may help explain why some lattice calculations find results consistent with larger symmetry groups than that dictated by large-\Nc alone \cite{Wagman:2017tmp,NPLQCD:2020lxg}.

\subsection{Including strangeness}
\label{sec:strange}

The analysis of  the two-flavor case can be generalized to three flavors.
The Savage-Wise coefficients of the octet-octet couplings \cite{Savage:1995kv} have previously been related to the parameters of the SU(6) Lagrangian \cite{Kaplan:1995yg}. 
While there is no analysis of the octet-decuplet and decuplet-decuplet couplings from the SU$(2 F)$ invariant Lagrangian available, there has been an analysis in chiral EFT \cite{Haidenbauer:2017sws}.
Because of phenomenological and lattice QCD work done on \Swave \OO interactions, we consider them here in the context of SU(2$F$) imposed on a pionless EFT including octet and decuplet baryons.
For \OO scattering in the two \Swave channels we find 
    \begin{align}
    \label{eq:COOspinzero}
         \COOoneS &= 2\left(\at - \frac{\bt}{3}\right) \\
    \label{Eq:COOspintwo}
        \COOfiveS &= 2\left(\at + \frac{\bt}{3}\right)  .
    \end{align}
Comparison with Eqs.~\eqref{eq:CDDabNNchannel:2} and \eqref{eq:CDDabNNchannel:4} shows that 
\begin{align}
    \label{eq:OODDEq1} \COOoneS &= \CDDoneSIthree \, ,\\ 
    \COOfiveS &= \CDDfiveSIthree \, ,
\end{align}
which is just a consequence of flavor SU(3) symmetry \cite{Haidenbauer:2017sws}. But it also shows that in the SU(6)-symmetric limit 
\begin{align}
    \label{eq:OODDEq2} \COOoneS & = \CDDsevenSIzero \, , \\
    \COOfiveS & = \CDDsevenSItwo \, .
\end{align}
In the notation of Ref.~\cite{Haidenbauer:2017sws}, 
\begin{align}
    C^{28,\oneS} &= C^{\overline{10},\sevenS} \, ,\\
    C^{28,\fiveS} &= C^{35,\sevenS} \, .
\end{align}
In addition, Eqs.~\eqref{eq:CDDabNNchannel:1} and \eqref{eq:CDDabNNchannel:3} give
\begin{align}
    C^{\overline{10},\threeS} & = C^{27,\oneS} \, , \\
    C^{27,\fiveS} &= C^{35,\threeS} \, . 
\end{align}
The equality of the LECs also yields the equality of the LO scattering amplitudes in the corresponding channels, i.e., $\calA_{\OO}^{(\oneS)} = \calA_{\DD}^{(0,3)} = \calA_{\DD}^{(3,0)}$ and $\calA_{\OO}^{(\fiveS)}$ = $\calA_{\DD}^{(2,3)} = \calA_{\DD}^{(3,2)}$.

Considering Eq.~\eqref{eq:OODDEq2}, for example, and the general form of the LECs shown in Eq.~\eqref{eq:LECgeneral}, we see that $\mu$-independent amplitudes require that $M_\Omega=M_\D$, and that either the scattering lengths in these channels are the same, i.e.,
\begin{align}
    \aOOsing =  a_{\DD}^{(3,0)} \, ,
\end{align}
or that they are each sufficiently large to be perturbatively close to the unitary limit, as is the physical case for \NN scattering.

In Ref.~\cite{Gongyo:2017fjb}, the HAL QCD collaboration argues that the \OO system in the $\oneS, I=0$ channel is close to the unitary limit.
(The earlier result of Ref.~\cite{Buchoff:2012ja} disagrees.)
Because of the equality with the $\mathscr{S}=3,I=0$ \DD channel, the same conclusions about the unitary limit hold:
    \begin{align}
        \at(\mu) \to -\frac{2 \pi}{M \mu}\, , \quad \bt(\mu)  \to 0\, . 
    \end{align}
This suggests that the octet and decuplet  sectors decouple.
The original Lagrangian was constructed to be SU(6) invariant with fields transforming under the 56-dimensional representation of SU(6), and this symmetry has now been enhanced to an SU(16) symmetry for octet baryons, SU(40) for decuplet baryons, or SU(56) for the octet and decuplet baryons combined.
Further, if the couplings sit exactly at the points above,  the \NN and \DD systems are also in the unitary limit.

\subsection{The role of entanglement}
\label{sec:entangle}

Before concluding, we speculate on a possible connection between the large-\Nc limit and the suppression of spin entanglement in baryon-baryon scattering. We point out possible future work that could help illuminate this potential relationship.
References~\cite{Beane:2018oxh,Low:2021ufv,Liu:2022grf,Bai:2023tey,Miller:2023ujx} consider spin entanglement in the scattering of spin-1/2 baryons at low enough energies that only $S$-waves are relevant using a variety of entanglement measures. 
In particular, Ref.~\cite{Beane:2018oxh} found that for the two-flavor case (i.e., \NN scattering), the spin entanglement power \cite{Zanardi:2001zza,Ballard:2011} is minimized when the \NN interactions possess a Wigner symmetry. 
Wigner symmetry also follows from the large-\Nc limit \cite{Kaplan:1995yg}, independent of the value of \bt. 
Spin entanglement suppression and the large-\Nc limit therefore lead to the same symmetry in \NN scattering.

For the three-flavor case with an SU(6) large-\Nc symmetry, the six independent octet-octet LECs \cite{Savage:1995kv} can be expressed in terms of \at and \bt as discussed above. 
However, only one LEC remains nonzero in the $\bt \to 0$ limit, resulting in an enlarged SU(16) symmetry in these interactions \cite{Kaplan:1995yg}. 
Reference \cite{Beane:2018oxh} showed that this SU(16) symmetry that arises from $\bt=0$ corresponds to a suppression of the spin entanglement power in octet-octet scattering.
Entanglement suppression therefore provides a more stringent constraint on the interactions of spin-1/2 baryons in the three-flavor sector than the large-\Nc limit, unlike in the two-flavor case. 
As shown above, $\bt=0$ also follows from the unitary limit for baryon-baryon scattering. A connection between the unitary limit and entanglement suppression was observed in Ref.~\cite{Beane:2018oxh}.

However, setting $\bt=0$ does have an effect in the two-flavor case when \D's are considered. 
It would be interesting to investigate whether entanglement suppression in \DD scattering also corresponds to an enlarged symmetry with $\bt=0$, thus being more constraining than the large-\Nc limit alone.\footnote{While this manuscript was under review, Ref.~\cite{Hu:2025lua} appeared.}

\section{Conclusion}

The large-\Nc expansion provides a framework for analyzing the relative importance of different contributions to the \NN interactions.
So far, the large-\Nc analysis has been applied to theories and models without explicit \D degrees of freedom.
In general the results of these analyses agree with data where available, despite the importance of the \D for the internal consistency of the large-\Nc approach applied to baryons.
In this paper we investigate possible explanations for why large-\Nc symmetry ordering applied to theories without a dynamical \D are largely consistent with available measurements.  Below we summarize our findings.

\begin{itemize}

\item For purely \Swave \NN interactions, operator relationships dictated by large-\Nc are unchanged whether a dynamical \D is included or not. 
These relationships are given as ratios of LECs, and as a consequence of the SU(4) symmetry that emerges in the large-\Nc limit the \D contributions cancel; a decoupling of the \D degree of freedom is not required.
The equalities that are required for this cancellation to occur are derived from a spin-isospin exchange symmetry that is valid for \Swave interactions.

\item In \D-less \eftnopi, \NN interactions in $P$ waves and most other higher partial waves are perturbative. 
Assuming that \ND and \DD interactions can similarly be treated perturbatively, \D contributions can only contribute to \NN observables at higher orders than \NN interactions. 
The situation is less clear for interactions in which higher partial waves couple to $S$ waves and for large-\Nc-based relationships involving $S$ and $P$ waves.
In these cases, the cancellations of the \D contributions seen in the purely \Swave sector are not evident. The results in \D-less \eftnopi emerge if the coupling of the \DD sector to the \NN sector vanishes or is perturbatively small.

\item The large-\Nc SU(4) symmetry imposes stringent constraints on the \NN, \DN, and \DD LECs. 
We derived the expressions of these LECs in terms of the two SU(4) parameters \at and \bt.
In principle, once these two parameters are determined from two \Swave scattering channels, the scattering amplitudes in the remaining channels can be related. 
Consistency requires that the baryon masses all be equal at LO in large-\Nc, and this is evident by renormalization invariance.

\item If the SU(2$F$) parameter $\bt=0$, the \NN and \DD sectors decouple. 
Apparently, this can be achieved either if one baryon-baryon scattering channel is in the unitary limit, which then drives others to their unitary limit, or if two scattering lengths that are not related to each other by the SU($2F$) symmetry are equal. 
We observe that the proximity of the physical \NN scattering lengths to the unitary limit drives \bt towards zero, resulting in the suppression of \DD contributions to \NN scattering and an enlarged symmetry of the theory, namely SU(20) for $F=2$ and SU(56) for $F=3$.

\item
Entanglement suppression is consistent with a vanishing \bt, but entanglement studies have not yet included the spin-3/2 particles in SU(2$F$).\footnotemark[\value{footnote}] 

\end{itemize}

\begin{acknowledgments}

RPS thanks Mark Wise for the initial ``flip" observation and subsequent discussions, and Hersh Singh for discussions.
This material is based upon work supported by the U.S.~Department of Energy, Office of Science, Office of Nuclear Physics,
under Award Numbers DE-SC0019647 (MRS) and DE-FG02-05ER41368 (TRR, RPS).
This work was supported in part by the Deutsche Forschungsgemeinschaft (DFG) through the Cluster of Excellence ``Precision Physics, Fundamental Interactions, and Structure of Matter'' (PRISMA${}^+$ EXC 2118/1) funded by the DFG within the German Excellence Strategy (Project ID 39083149), by the NSF through cooperative agreement 2020275, and by the DOE Topical Collaboration “Nuclear Theory for New Physics,” award No. DE-SC0023663 (TRR).

\end{acknowledgments}

\appendix
\section{\DD partial-wave projections}
\label{sec:AppLagrangians}

Here we present details of the \Swave Lagrangians containing \D fields. The notational conventions differ from those of Ref.~\cite{Savage:1996tb}. 
The $\D N$ Lagrangian is given in Eq.~\eqref{eq:NDLagShort}, while the \DD Lagrangian is given in very compact notation in Eq.~\eqref{eq:DDLagShort}.
The $\mathscr{P}^{(\mathscr{S},I)}$ are operators that project onto \DD $S$-wave channels with total spin $\mathscr{S}$ and isospin $I$. 
To avoid an overly cluttered presentation, the notation used for the \D bilinears corresponds to
\begin{align} \label{eq:DDProjectorShorthand}
    \D^T \mathscr{P}^{(\mathscr{S},I)} \D \equiv \D_{\alpha\beta\gamma}^{abc}    \left[\mathscr{P}^{(\mathscr{S},I)}\right]^{\alpha\alpha',\beta\beta',\gamma\gamma'}_{aa',bb',cc'}
    \D_{\alpha'\beta'\gamma'}^{a'b'c'} \, .
\end{align}
Depending on the values of $\mathscr{S}$ and $I$, the operators $\mathscr{P}^{(\mathscr{S},I)}$ carry different numbers of spin ($K,L,\ldots$) and/or isospin ($A,B,\ldots$) indices with $\{A,B,\ldots,K,L,\ldots\} = 1,2,3$. 
For example, the operator projecting onto the $\sevenS, I=2$ channel is $\mathscr{P}^{(3,2)}_{KLM,AB}$.
In Eq.~\eqref{eq:DDLagShort}, all spin and isospin indices are contracted between the two bilinears to make rotational and isospin scalars, just as they are in the simpler case of Eq.~\eqref{eq:NNLag}.

Using the notation that parentheses around spin and isospin matrices denote the first combination of indices (i.e., ${}_{aa'}^{\alpha\alpha'}$ in Eq.~\eqref{eq:DDProjectorShorthand}), angled brackets the second, and curly brackets the third combination, the projection operators are given by 
\begin{align}
\mathscr{P}^{(0,1)}_A & = 
        \frac{3}{4\sqrt{10}}(\sigma_2\tau_2\tau_A)\langle\sigma_2\tau_2\rangle \{\sigma_2\tau_2 \} \, , \label{eq:P1S0I1} \\
\mathscr{P}^{(0,3)}_{ABC} & = 
        \frac{1}{8} \left[ \delta_{AD}\delta_{BE}\delta_{CF} - \frac{1}{5}\delta_{EF}(\delta_{AB}\delta_{CD}+\delta_{AC}\delta_{BD}+\delta_{BC}\delta_{AD}) \right] 
        (\sigma_2\tau_2\tau_D)\langle\sigma_2\tau_2\tau_E\rangle \{\sigma_2\tau_2\tau_F \} \, , \nonumber \\
    &= 
        \frac{1}{8} \left[ (\sigma_2\tau_2\tau_A)\langle\sigma_2\tau_2\tau_B\rangle \{\sigma_2\tau_2\tau_C \} 
        -\frac{1}{5} \delta_{AB}(\sigma_2\tau_2\tau_C)\langle\sigma_2\tau_2\rangle \{\sigma_2\tau_2 \} \right. \\
    &\mathrel{\phantom{=}} \left.   -\frac{1}{5}\delta_{AC}(\sigma_2\tau_2\tau_B)\langle\sigma_2\tau_2\rangle \{\sigma_2\tau_2 \} 
        -\frac{1}{5}\delta_{BC}(\sigma_2\tau_2\tau_A)\langle\sigma_2\tau_2\rangle \{\sigma_2\tau_2 \}\right]\, , \nonumber \\
\mathscr{P}^{(1,0)}_K & = 
        \frac{3}{4\sqrt{10}} (\sigma_2\sigma_K\tau_2)\langle\sigma_2\tau_2\rangle \{\sigma_2\tau_2 \} \, , \\
\mathscr{P}^{(1,2)}_{K,AB} & = 
        \frac{3\sqrt{3}}{8\sqrt{5}} \left[ \delta_{AD}\delta_{BE} - \frac{1}{3}\delta_{DE} \right]
        (\sigma_2\sigma_K\tau_2\tau_D)\langle\sigma_2\tau_2\tau_E\rangle \{\sigma_2\tau_2 \} \, , \\
\mathscr{P}^{(2,1)}_{KL,A} & =  
        \frac{3\sqrt{3}}{8\sqrt{5}} \left[ \delta_{KX}\delta_{LY} - \frac{1}{3}\delta_{XY} \right](\sigma_2\sigma_X\tau_2\tau_A)\langle\sigma_2\sigma_Y\tau_2\rangle \{\sigma_2\tau_2 \} \, ,\\
\mathscr{P}^{(2,3)}_{KL,ABC} & =  
        \frac{\sqrt{3}}{8\sqrt{2}} \left[ \delta_{KX}\delta_{LY} - \frac{1}{3}\delta_{XY} \right]
        \left[ \delta_{AD}\delta_{BE}\delta_{CF} - \frac{1}{5}\delta_{EF}(\delta_{AB}\delta_{CD}+\delta_{AC}\delta_{BD}+\delta_{BC}\delta_{AD}) \right] \nonumber\\
    &\mathrel{\phantom{=}}(\sigma_2\sigma_X\tau_2\tau_D)\langle\sigma_2\sigma_Y\tau_2\tau_E\rangle \{\sigma_2\tau_2\tau_F \} \, ,\\
\mathscr{P}^{(3,0)}_{KLM} & = 
        \frac{1}{8} \left[ \delta_{KX}\delta_{LY}\delta_{MZ} - \frac{1}{5}\delta_{YZ}(\delta_{KL}\delta_{MX}+\delta_{KM}\delta_{LX}+\delta_{LM}\delta_{KX}) \right] 
        (\sigma_2\sigma_X\tau_2)\langle\sigma_2\sigma_Y\tau_2\rangle \{\sigma_2\sigma_Z\tau_2 \} \, , \\
\mathscr{P}^{(3,2)}_{KLM,AB} & =  
        \frac{\sqrt{3}}{8\sqrt{2}} 
        \left[ \delta_{KX}\delta_{LY}\delta_{MZ} - \frac{1}{5}\delta_{YZ}(\delta_{KL}\delta_{MX}+\delta_{KM}\delta_{LX}+\delta_{LM}\delta_{KX}) \right]  \left[ \delta_{AD}\delta_{BE} - \frac{1}{3}\delta_{DE} \right]\nonumber\\
    &\mathrel{\phantom{=}}(\sigma_2\sigma_X\tau_2\tau_D)\langle\sigma_2\sigma_Y\tau_2\tau_E\rangle \{\sigma_2\sigma_Z\tau_2 \} \, \label{eq:P7S3I2} .
\end{align}

The $\D^{abc}_{\alpha\beta\gamma}$ fields have a high degree of symmetry under index interchange; they are independently symmetric in the spin and in the isospin indices. The anticommutator of two \D fields is given by
\begin{align}
    \{\D^{abc}_{\alpha\beta\gamma},(\D^\dagger)^{a'b'c'}_{\alpha'\beta'\gamma'} \} = \frac{1}{6}\left[\delta_{aa'}\delta_{bb'}\delta_{cc'} + \text{(all perm.)}\right] \frac{1}{6}\left[\delta_{\alpha\alpha'}\delta_{\beta\beta'}\delta_{\gamma\gamma'} + \text{(all perm.)} \right] \, ,
\end{align}
where (all perm.) denotes all possible permutations of the indices. This ensures that the physical \D fields obey canonical anticommutation relations.
The normalization of the operators in Eqs.~\eqref{eq:P1S0I1}-\eqref{eq:P7S3I2} is chosen such that each addition of a \DD loop in a scattering diagram corresponds to a factor of $C_{\DD}^{(\mathscr{S},I)} I^{\DD}$.

\bibliography{biblio}

\end{document}